\documentclass[11pt,english]{article}
\usepackage{amsfonts, amsmath, amssymb, verbatim, setspace, pdfsync, graphicx,bigints, mathtools}
\usepackage{comment}
\usepackage{color}
\usepackage{pdfsync}
\definecolor{darkgreen}{rgb}{0,0.5,0}
\definecolor{darkblue}{rgb}{0,0,0.6}
\definecolor{purple}{rgb}{0.4,0.15,0.21}
\definecolor{black}{rgb}{.2,.2,.2}
\usepackage[colorlinks=true,citecolor=darkblue,linkcolor=black,urlcolor=darkblue]{hyperref}

\makeatletter
\newcommand*{\defeq}{\mathrel{\rlap{%
                     \raisebox{0.3ex}{$\m@th\cdot$}}%
                     \raisebox{-0.3ex}{$\m@th\cdot$}}%
                     =}
\makeatother

\DeclareMathOperator{\p}{\partial}
\DeclareMathOperator{\h}{\theta}
\DeclareMathOperator{\Tr}{Tr}

\newcommand{\be}{\begin{equation}}
\newcommand{\ee}{\end{equation}}

\newcommand{\f}{\frac}

\begin{document}
\unitlength = 1mm
\ 

\begin{center}

{ \LARGE {\textsc{\begin{center}Modular forms and a generalized Cardy formula in higher dimensions \end{center}}}}

\vspace{0.8cm}
Edgar Shaghoulian

\vspace{.5cm}

{\it Department of Physics} \\
{\it University of California}\\
{\it Santa Barbara, CA 93106 USA}

\vspace{1.0cm}

\end{center}

\begin{abstract}

\noindent We derive a formula which applies to conformal field theories on a spatial torus and gives the asymptotic density of states solely in terms of the vacuum energy on a parallel plate geometry. The formula follows immediately from global scale and Lorentz invariance, but to our knowledge has not previously been made explicit. It can also be understood from the fact that $\log Z$ on $\mathbb{T}^2\times \mathbb{R}^{d-1}$ transforms as the absolute value of a non-holomorphic modular form of weight $d-1$, which we show. The results are extended to theories which violate Lorentz invariance and hyperscaling but maintain a scaling symmetry. The formula is checked for the cases of a free scalar, free Maxwell gauge field, and free $\mathcal{N}=4$ super Yang-Mills. The  case of a Maxwell gauge field gives Casimir's original calculation of the electromagnetic force between parallel plates in terms of the entropy of a photon gas.

\end{abstract}

\pagebreak
\setcounter{page}{1}
\pagestyle{plain}

\setcounter{tocdepth}{1}


\section{Introduction}\label{intro}
The study of two-dimensional conformal field theory (CFT) contains many rich and powerful results, with wide applications from condensed matter systems to quantum gravity and holography. One of the central tools of the theory  is modular invariance on a Euclidean torus background. Among other things,  modular invariance implies a duality between the partition function at high temperature and at low temperature. Such dualities often lead to strong analytic results. As a primitive example, the high-temperature/low-temperature duality of Kramers and Wannier exactly determined the critical point of the two-dimensional Ising model \cite{Kramers:1941kn}. In the case of a generic conformal field theory, the temperature-inversion duality can be used to derive the Cardy formula, which depends only on the central charge of the CFT and gives the degeneracy of states at high energy in the Hilbert space \cite{Cardy:1986ie}. The states are on a circle and are in one-to-one correspondence with local operators on $\mathbb{R}^2$. The derivation of the formula uses two key facts of two-dimensional CFTs. The first is the high-temperature/low-temperature duality already mentioned, which allows one to project the high-temperature partition function to the contribution of the vacuum sector of the theory. Second, the vacuum energy (or Casimir energy) is provided by the anomalous transformation of the stress-energy tensor when conformally mapping the plane to the unit cylinder, $E_{\textrm{vac}}=-c/12$.
 
It is natural to wonder about higher-dimensional analogs of modular invariance, temperature-inversion dualities, and the Cardy formula \cite{Cardy:1991kr}. The primary difficulty lies in the possibility of spatial curvature. To count states which are in one-to-one correspondence with local operators, one has to consider the manifold $S^d \times \mathbb{R}$, where the $S^d$ is curved for $d>1$. This curvature couples to the CFT fields and spoils the possibility of a universal formula. Furthermore, the spatial curvature introduces scheme-dependence into the calculation of the Casimir energy through possible counterterms which are integrals of local curvature invariants \cite{Lorenzen:2014pna, Assel:2015nca}.\footnote{Interestingly, strongly coupled field theories with holographic duals seem to admit a formula on $S^d \times \mathbb{R}$, known as the Cardy-Verlinde formula \cite{Verlinde:2000wg}. This formula is only universal for holographic CFTs and becomes ambiguous at weak coupling \cite{Kutasov:2000td}. Some interesting higher-dimensional supersymmetric Cardy formulas have recently been constructed in \cite{DiPietro:2014bca, Ardehali:2015hya, Zhou:2015cpa}, and an extension of the parity-odd part of the Cardy formula to higher dimensions \cite{Landsteiner:2011cp, Landsteiner:2011iq, Loganayagam:2012pz, Loganayagam:2012zg, Jensen:2012kj, Jensen:2013kka, Jensen:2013rga} is reviewed in \cite{Azeyanagi:2015gqa}. In two dimensions, there are extensions of the formula to non-conformal field theories \cite{Detournay:2012pc, Bagchi:2012xr, Shaghoulian:2015dwa}.}

In this paper  we will instead consider an arbitrary CFT on $\mathbb{T}^d \times \mathbb{R}$. This manifold is not related to $\mathbb{R}^{d+1}$ by a conformal transformation unless $d=1$. At finite temperature the thermal partition function should be invariant under $SL(d+1,\mathbb{Z})$, which contains the invariance of swapping cycles. We will see that the Cardy formula naturally generalizes and gives formulas for the thermal entropy at high temperature and asymptotic density of states in terms of the vacuum (or Casimir) energy of the CFT on $S^1\times \mathbb{R}^{d-1}$:
\begin{align}
S &= (d+1)T^d V_d \varepsilon_{\textrm{vac}}\,,\nonumber\\
\log \rho(E) &= \f{d+1}{d^{\f{d}{d+1}}}\left(\varepsilon_{\textrm{vac}}V_d\right)^{\f{1}{d+1}}E^{\f{d}{d+1}}\,.
\end{align}
This does not really require conformal invariance. As we will review below, for any relativistic QFT, the free energy density and the Casimir energy density are two sides of the same coin. Adding scale invariance to the mix then shows that the thermal entropy is governed entirely by a scale-independent number $\varepsilon_{\textrm{vac}}$ which characterizes the Casimir energy. This dimensionless  number is not given by the anomalies of the theory and is generically dependent on the coupling constants. Whereas the partition function in two dimensions is a modular invariant function, we will find that the logarithm of the partition function transforms as the absolute value of a (not necessarily holomorphic) modular form of weight $d-1$:
\be
\log Z\left(\f{a\tau+b}{c\tau+e},\f{a\bar{\tau}+b}{c\bar{\tau}+e}\right)= \big|(c\tau+e)^{d-1}\big|\log Z(\tau, \bar{\tau})  \,,
\ee
where $\tau$ is the modular parameter of the two-torus made up of the thermal cycle and finite spatial cycle. To be clear, the modular invariance is not anomalous; the reason for the prefactor is that we have kept the size of the spatial background fixed under modular transformations that would otherwise change it. This is enforced by using the usual two-dimensional modular parameter $\tau$. The logarithm of the partition function is still invariant under a general $SL(d+1,\mathbb{Z})/\mathbb{Z}_2$ transformation. This will become clear in the next section. We can also define a modular-invariant function 
\be
F(\tau,\bar{\tau}) \defeq \beta^{\f{d-1}{2}}\log Z(\tau,\bar{\tau}) = F\left(\f{a\tau+b}{c\tau+e},\f{a\bar{\tau}+b}{c\bar{\tau}+e}\right)
\ee
with $\beta = \textrm{Im } \tau$. Alternatively, we could have defined an invariant density by dividing by the spatial volume.

These formulas induce a high-temperature/low-temperature duality from which the entropy formulas descend. We will also derive analogous formulas for theories which violate Lorentz invariance and hyperscaling but maintain an anisotropic scaling symmetry. 

\section{Swapping cycles}\label{genmodular}
We will begin with a quantum theory defined by a Euclidean path integral. The theory will be on a rectangular spatial torus $\mathbb{T}^d$ with no twists, i.e. no angular potentials. We will specify the necessary symmetries as we proceed. We will not actually  use the full conformal symmetry except when we comment on the connection to local operators below. 

The partition function of the theory at inverse temperature $\beta$ may be given as 
\be
Z(\beta) = \int [\mathcal{D}\Phi]\;e^{-I_E} = \Tr_{L\times \mathbb{T}^{d-1}} e^{-\beta H} \,.
\ee
The field $\Phi$ is a general placeholder for the fields of the theory. The spatial manifold on which the Hilbert space is defined is written explicitly as $L\times \mathbb{T}^{d-1}$, i.e. one of the cycles has length $L$.  Using Euclidean rotational invariance between the thermal cycle and the spatial cycle of length $L$ to perform a 90 degree rotation, we can write the partition function as
\be
Z(\beta) = \Tr_{\beta\times \mathbb{T}^{d-1}} e^{-L H}\,.
\ee 
To admit a correct interpretation as a thermal partition function, one needs to assign the usual thermal periodicity conditions along the cycle $L$. 

We now take $L$  large to project the partition function to the ground state of the theory on the  torus $\beta\times \mathbb{T}^{d-1}$. This gives 
\be\label{swapcycle}
Z(\beta) \approx e^{-L E_{\textrm{vac}, \,\beta\times \mathbb{T}^{d-1}}}\,.
\ee
Finally, we take $\beta$ smaller than all the other cycles of the torus and assume the theory is scale invariant. In that case, the background is effectively $S^1 \times \mathbb{R}^{d-1}$ with periodicity $\beta$ for the $S^1$, so the vacuum energy on this background is given by scale invariance as $E_{\textrm{vac}, \,\beta\times \mathbb{T}^{d-1}} = -\varepsilon_{\textrm{vac}} V_{\beta\times \mathbb{T}^{d-1}}/\beta^{d+1}$. $V_{\beta\times \mathbb{T}^{d-1}}$ is the volume of the  spatial torus $\beta \times \mathbb{T}^{d-1}$ (which is effectively $S^1 \times \mathbb{R}^{d-1}$) and $\varepsilon_{\textrm{vac}}$ is a pure number independent of any dimensionful scales. The spatial $S^1$ inherits thermal periodicity conditions. This gives 
\be\label{torusswap}
\log Z(\beta) \approx \varepsilon_{\textrm{vac}} L V_{\beta\times \mathbb{T}^{d-1}}/\beta^{d+1}= \varepsilon_{\textrm{vac}} V_{L\times \mathbb{T}^{d-1}}/\beta^d\,,
\ee
where in the final expression we translated to quantities for the original torus using $LV_{\beta\times \mathbb{T}^{d-1}} =\beta V_{L\times \mathbb{T}^{d-1}} $. We can now get the thermodynamic entropy in terms of the  volume of the original spatial torus $V_d :=  V_{L\times \mathbb{T}^{d-1}}$ as 
\be\label{formula}
S =(1-\beta \partial_{\beta})\log Z= (d+1) T^d V_d \,\varepsilon_{\textrm{vac}}\,.
\ee
The microcanonical density of states is obtained by inverse Laplace transforming the partition function:
\be\label{states}
\rho(E) = \int d\beta \,Z(\beta)\, e^{\beta E}\implies \log \rho(E)=\f{d+1}{d^{\f{d}{d+1}}}\; (\varepsilon_{\textrm{vac}} V_d)^{\f{1}{d+1}} E^{\f{d}{d+1}}\,,
\ee
where we evaluated the integral by a saddle-point approximation for large $E$, obtaining $\beta^*=(d\,\varepsilon_{\textrm{vac}} V_d/E)^{\f{1}{d+1}}$. This shows that the thermodynamic entropy at large $T$ and the degeneracy of states at large $E$ are given solely in terms of the Casimir energy at zero temperature on $S^1 \times \mathbb{R}^{d-1}$, where the periodicity conditions along the $S^1$ are inherited from the periodicity conditions along the original thermal cycle.

It should be noted that $\varepsilon_{\textrm{vac}}$ is not related to the anomalies of the theory. This is easily illustrated by the example of $\mathcal{N}=4$ super Yang-Mills, for which $\varepsilon_{\textrm{vac}}$ depends on the exactly marginal `t Hooft coupling. For a CFT at high temperature we can often ignore the effects of spatial curvature, so we can consider the spatial manifold to be a sphere. By the state-operator correspondence this could equally well be a formula for the degeneracy of local operators of scaling dimension $E$ of the CFT.

\subsection{Modular invariance I}\label{mod1}
So far we have only used the relationship between high and low temperatures asymptotically. To precisely mimic what happens in $1+1$ dimensions (i.e. a high-temperature/low-temperature duality where a given CFT on a given background at a given temperature maps to the same CFT on the same background at some inversely related temperature), we can consider the spatial manifold $S^1 \times \mathbb{R}^{d-1}$. We pick periodicity $L$ for the $S^1$. The plane should be understood as regulated into a large torus with equal-length cycles $L_\infty\gg \{\beta, \,L, \,L^2/\beta\}$. We will continue to write sums for the partition function since the spectrum can be discretized in this way. 

We swap the thermal and spatial $S^1$ cycles and perform a scale transformation by $L/\beta$. This scale transformation restores the spatial cycle to size $L$ and changes the thermal cycle to size $L^2/\beta$. The volume of the new plane $\widetilde{\mathbb{R}}^{d-1}$ has increased by a factor of $(L/\beta)^{d-1}$. This relates the partition function on the background $S^1\times\mathbb{R}^{d-1}$ to the partition function on the larger background $S^1\times \widetilde{\mathbb{R}}^{d-1}$:
\be
Z(\beta) = \sum e^{-\beta E} = \sum e^{-(L^2/\beta) \widetilde{E}}=\widetilde{Z}\left(\f{L^2}{\beta}\right)\,.
\ee
Due to the large volume of the plane, we expect that $\log Z \sim V_{d-1}$. We can therefore scale out the factor of $(L/\beta)^{d-1}$ from the volume of the larger plane to return to our original plane. This gives the following high-temperature/low-temperature duality in $d+1$ dimensions:
\be
\log Z(\beta) = \left(\f{L}{\beta}\right)^{d-1}\log Z\left(\f{L^2}{\beta}\right)\,.
\ee
This relation applies at any $\beta$ and $L$, and it has corrections suppressed by inverse powers of the length scale $L_\infty$. The self-dual point of the transformation is $\beta = L$. To end with the same theory after swapping cycles, one should either start with the same periodicity conditions along both cycles or sum over all possible structures. The entropy can be derived directly from this invariance and agrees with that of the previous subsection. There is a subtlety associated with the fact that there are many states close to the vacuum; see appendix \ref{app} for details.

Defining $\tau$ as the usual modular parameter of the two-dimensional torus will let us make a statement about the general $SL(2,\mathbb{Z})$ transformation
\be
\tau \to \f{a\tau+b}{c\tau+e}\,, \qquad \bar{\tau} \to \f{a\bar{\tau}+b}{c\bar{\tau}+e}\,.\label{sl2z}
\ee
This requires an overall rigid rescaling of the torus by $|c\tau+e|^{-1}$. Performing the same trick as before, we scale out this factor from the volume of $\widetilde{\mathbb{R}}^{d-1}$ to get
\be
\log Z(\tau, \bar{\tau}) = \big|(c\tau+e)^{1-d}\big| \log Z\left(\f{a\tau+b}{c\tau+e},\f{a\bar{\tau}+b}{c\bar{\tau}+e}\right)\,.
\ee
This implies that $\log Z(\tau, \bar{\tau})$ transforms as the absolute value of a (not necessarily holomorphic) modular form of weight $d-1$. Holomorphy would require Im$[\tau] \rightarrow \textrm{Im}[\tau]+i$ as a symmetry, which generally fails as in the case of free field theories with $d>1$.

We can transform this into a modular-invariant function $F(\tau,\bar{\tau})$ by using  Im$\left[\f{a\tau+b}{c\tau+e}\right]=\textrm{Im}[\tau]/|c\tau+e|^2$: 
\be
F(\tau,\bar{\tau}) \defeq \textrm{Im}[\tau]^{\f{d-1}{2}} \log Z( \tau,\bar{\tau}) = F\left(\f{a\tau+b}{c\tau+e}, \f{a\bar{\tau}+b}{c\bar{\tau}+e}\right) \,.
\ee
In more physical notation, and ignoring angular momentum, we have
\be
F(\beta) = \beta^{\f{d-1}{2}} \log Z(\beta) = F\left(\f{L^2}{\beta}\right)\,.
\ee
\vspace{-10mm}

\subsection{Modular invariance II}\label{mod2}
There is another special torus on which a high-temperature/low-temperature duality becomes transparent. It is the torus with cycle lengths $\beta$, $L$, and $L_{i} = (L/\beta)L_{i-1}$ for $i=1,\dots,d-1$ with $L_0\defeq L$. In this case, we can swap the $\beta$ and $L_{d-1}$ cycles and rescale by $L/\beta$. The spatial cycle lengths of the torus under this combined rotation and scaling remain invariant but become permuted among one another. They can be taken back to their original orientation by $d-1$ swaps of cycle pairs. This leaves us with precisely the same spatial torus, with a thermal cycle of length $L_{d-1} L/\beta = L (L/\beta)^d$. In other words, 
\be
Z(\beta) = Z\left(L\left(\f{L}{\beta}\right)^d\right)\label{mod2eq}
\ee
on this special torus. This invariance differs from the one in the previous subsection, but mimics the invariance of the two-dimensional hyperscaling-violating theories discussed in section \ref{generalizations}. As a consistency check, notice that the thermal entropy at large temperature obtained from this invariance is the same as derived in the previous subsections. 
\vspace{-1mm}
\subsection{Modular invariance I$+$II}\label{mod3}
We can combine the methods of the previous two subsections by taking a torus with $n$ of the directions combining into the special torus of section \ref{mod2} and the other $d-n$ directions are large. This gives
\be
\log Z(\beta) = \left(\f{L}{\beta}\right)^{d-n}\log Z\left(L\left(\f{L}{\beta}\right)^n\right).
\ee
\section{Generalizations}\label{generalizations}
\subsection{Hyperscaling-violation}
We first consider a two-dimensional theory with hyperscaling-violation exponent $\h$ on $\mathbb{T}^2$. By hyperscaling violation we mean that the stress-energy tensor has dimension $2-\h$: $T_{\mu \nu}(\lambda x) = \lambda^{-2+\h} T_{\mu \nu}(x)$.\footnote{In holographic models of hyperscaling violation the shift $\h$ is more appropriately associated with a scaling dimension of the quantum state, but since the partition function under consideration here only involves the one-point function $\langle T_{tt} \rangle$ this will make no difference for us.} After swapping cycles we can write a trivial equivalence as below:
\be
Z(\beta) = \sum e^{-\beta \int_0^L dx\, T_{tt}(x)}\underbrace{=}_{\textrm{swap}} \sum e^{-L \int_0^\beta dx\, T_{tt}(x)} \underbrace{=}_{\textrm{trivial}} \sum e^{-\lambda^{1-\h} L \int_0^\beta \lambda \,dx \lambda ^{-2+\h}T_{tt}(x)}. 
\ee
Upon identifying $\lambda = L/\beta$ we can  reinterpret the integral in the final expression as the energy on length $L$, which leaves us with an invariance of the form 
\be
Z(\beta) = Z\left(L\left(\f{L}{\beta}\right)^{1-\h}\right)\,.
\ee
This is the same type of invariance as \eqref{mod2eq}. In this case, however, we can define a ``refined" partition function which is modular invariant in the usual sense:
\be
Z_{\textrm{ref}}(\beta) \defeq \sum e^{-\beta E^{1/(1-\h)}} \implies Z_{\textrm{ref}}(\beta) = Z_{\textrm{ref}}\left(\f{L^2}{\beta}\right).
\ee
In higher dimensions, the special torus of section \ref{mod2} gives
\be
Z(\beta) = Z\left(L\left(\f{L}{\beta}\right)^{d-\h}\right)\,.
\ee
We can also consider the background $\mathbb{T}^2 \times \mathbb{R}^{d-1}$ and use the results of section \ref{mod1} to get
\be
\log Z(\beta) = \left(\f{L}{\beta}\right)^{d-1} \log Z\left(L\left(\f{L}{\beta}\right)^{1-\h}\right)\,.
\ee
Using the refined partition function defined above, we have
\be
\log Z_{\textrm{ref}}(\beta) = \left(\f{L}{\beta}\right)^{d-1}\log Z_{\textrm{ref}}\left(\f{L^2}{\beta}\right)\,.
\ee
From the refined partition function we can define a modular-invariant function as before:
\be
F_{\textrm{ref}}(\tau, \bar{\tau}) \defeq \textrm{Im}[\tau]^{\f{d-1}{2}}\log Z_{\textrm{ref}}(\tau, \bar{\tau})\,.
\ee
The asymptotic density of states follows from these modular properties, but we will obtain it for a more general class of theories in the next subsection. 

\subsection{Anisotropic scaling}
We now consider a theory which has an anisotropic scaling symmetry and which violates hyperscaling. We stick to $\mathbb{T}^2\times\mathbb{R}^{d-1}$ and assume a rotational invariance between the cycles of the $\mathbb{T}^2$. A special torus analogous to section \ref{mod2} can be defined in this case as well, but we omit the details.

We begin by using rotational invariance to exchange the thermal cycle $\beta$ with the spatial cycle $L$. We can now perform the scaling symmetry $t\rightarrow \lambda\, t$, $x_i\rightarrow \lambda^{z_i}x_i$ with $x_1$ the coordinate along $L$ and $z_1=1$.  We pick $\lambda = L/\beta$ to restore the size of the spatial circle to $L$. This scales the new thermal cycle $L$ by a factor of $L/\beta$. The energy will transform anomalously according to the hyperscaling-violation exponent $\h$, giving a factor of $(L/\beta)^{-\h}$. The volume of the $\mathbb{R}^{d-1}$ piece rescales by $(L/\beta)^{\sum_{i=2}^d z_i}$, which we factor out as before. Altogether we have 
\be
\log Z(\beta) =\left(\f{L}{\beta}\right)^{\sum_{i=2}^d z_i}\log Z\left(L\left(\f{L}{\beta}\right)^{1-\h} \right)\,.
\ee
Defining $d_{\textrm{eff}} = \sum_{i=1}^d z_i-\h>0$, the high-temperature partition function projects to the ground state:
\be
Z(\beta) \approx \exp\left(-L^{d_{\textrm{eff}}+1}T^{d_{\textrm{eff}}}E_{\textrm{vac}}\right) \,
\ee
\be\label{thermogen}
\implies S =- (d_{\textrm{eff}}+1)  \,L^{d_{\textrm{eff}}+1}\,T^{d_{\textrm{eff}}} \,E_{\textrm{vac}}\,.
\ee
The vacuum energy scales as $E_{\textrm{vac}} \sim V_d/L^{d_{\textrm{eff}}+1}$ (notice that in these theories $[L_i] = -z_i$). There is an additional dimensionful scale in hyperscaling-violating theories, e.g. the Fermi momentum, which we have left out but can be inserted to restore dimensions. The asymptotic density of states is given as
\be\label{microgen}
\log \rho(E) = \f{d_{\textrm{eff}}+1}{d_{\textrm{eff}}\,^{\f{d_{\textrm{eff}}}{d_{\textrm{eff}}+1}}}\; (-E_{\textrm{vac}})^{\f{1}{d_{\textrm{eff}}+1}} E^{\f{d_{\textrm{eff}}}{d_{\textrm{eff}}+1}}L\,.
\ee
For $d=1$, $\h\neq 0$, and $z_i = 1$ this reduces to the formula in \cite{Shaghoulian:2015dwa, Bravo-Gaete:2015wua}.

Let us remark on theories with rotational invariance but without scale invariance, which may shed some light on the previous derivations. We begin with \eqref{swapcycle} to find 
\be\label{general}
S = (1-\beta \partial_\beta)\log Z = (d+1)T^d V_d\, \varepsilon_{\textrm{vac},\, \beta \times \mathbb{T}^{d-1}} - V_d\, T^{d-1}  \p_\beta \varepsilon_{\textrm{vac},\, \beta \times \mathbb{T}^{d-1}} \,.
\ee
We have again defined $E_{\textrm{vac},\, \beta \times \mathbb{T}^{d-1}} \defeq - \varepsilon_{\textrm{vac},\, \beta \times \mathbb{T}^{d-1}}V_{\beta \times \mathbb{T}^{d-1}}/\beta^{d+1}= - \varepsilon_{\textrm{vac},\, \beta \times \mathbb{T}^{d-1}} V_d/\beta^d$, where this time $\varepsilon_{\textrm{vac},\, \beta \times \mathbb{T}^{d-1}}$ can depend on the cycle lengths. We see that the benefits of a scaling symmetry were to (a) keep the background space the same and (b) determine the scaling of $E_{\textrm{vac}}$ with the dimensionful parameters of the theory.

\section{Examples}
In this section we perform a few checks of formula \eqref{formula}. It is important to note that when calculating Casimir energies, one only keeps pieces that have a dependence on the size of the $S^1$ and cannot be mimicked by the addition of a cosmological constant term to the action. Such pieces are observable.
\subsection{Free massless scalar}
The Casimir energy for a free massless scalar on $S^1 \times \mathbb{R}^{d-1}$  has been computed in \cite{Cappelli:1988vw}. The result for the energy is
\be 
\varepsilon_{\textrm{vac}} = \f{ \Gamma\left(\f{d+1}{2}\right)\zeta(d+1)}{\pi^{(d+1)/2}}\,.
\ee 
The thermal entropy of a free scalar gas in volume $V_d$ is given as (see e.g. \cite{Solodukhin:2011gn})
\be
S=\f{d+1}{\pi^{(d+1)/2}} \Gamma\left(\f{d+1}{2}\right)\zeta(d+1) T^d\, V_d\,.
\ee
We see that the thermal entropy and Casimir energy are related by formula \eqref{formula}. 

\subsubsection{Massive scalar}
We reviewed above that the equivalence of the free energy density with the Casimir energy density follows immediately from the Euclidean path integral. Let us then consider the massive scalar in four spacetime dimensions. We will only consider the leading correction due to the mass. The first correction to the Casimir energy density is \cite{Ambjorn:1981xw}
\be
\f{E_{\textrm{vac}}}{V} =-\f{\pi^2}{90 L^4}+ \f{m^2}{24 L^2}\,,\label{vacmass}
\ee
while the free energy density is
\be
\f{F}{V}=-\f{\pi^2}{90\beta^4}+\f{m^2}{24\beta^2}\,.\label{freemass}
\ee
These two are clearly the same expression up to interchanging $\beta\leftrightarrow L$. Obtaining the entropy from the free energy density will show that the leading pieces are related by \eqref{formula} whereas the subleading pieces are not. The more general formula \eqref{general} will relate the subleading pieces as well.

\subsection{Free Maxwell gauge field and the force between parallel plates}
The only difference in the Casimir energy and thermal entropy of a photon as compared to a scalar is a factor of $d-1$ to account for the number of polarization states of a massless gauge field in $d+1$ dimensions. Formula \eqref{formula} therefore accounts for this case as well. 

An interesting case is that of $d=3$, where we would like to compare to Casimir's original calculation of the induced pressure acting on parallel conducting plates in an electromagnetic medium \cite{Casimir:1948dh}. Tacking on the two polarization states of the photon to our massless scalar calculation, we have 
\be
\varepsilon_{\textrm{vac}}= 2\,\f{\zeta(4)\Gamma(2)}{ \pi^{2}}=\f{\pi^2}{45 }\,.
\ee
The entropy of a photon gas is given as 
\be
S=\f{4\pi^2}{45} V_3 \,T^3\,.
\ee
As already stated, these two expressions are connected by \eqref{formula}. 

Casimir's original calculation used Dirichlet boundary conditions at the surfaces of the plates, whereas \eqref{formula} relates the entropy of the photon gas to the vacuum energy on a torus, i.e. periodic boundary conditions. We can translate boundary conditions from periodic to Dirichlet by taking $L\rightarrow 2L$ in $\varepsilon_{\textrm{vac}} L^{-4}$ \cite{Ambjorn:1981xw, Brown:1969na}. 

In other words, we can derive Casimir's result by calculating the thermal entropy of a photon gas and deducing the Casimir energy on the torus from \eqref{formula}. Performing the rescaling $L\rightarrow 2L$ necessary to change boundary conditions gives
\be
E_{\textrm{vac}}^{QED} =\f{S}{64 T^3 L^4}= -\f{\pi^2 V_3}{720 L^4}\,.
\ee
Dividing by the volume and multiplying by $L$ gives the energy per unit area of the plates. Taking a derivative with respect to $L$ gives the pressure on the plates in terms of the thermal entropy of the photon gas:
\be
\f{F}{A} =-\p_L \left(\f{L\,E_{\textrm{vac}}^{QED}}{V_3}\right)=-\f{3\,S}{64 \,V_3 T^3 L^4}= -\f{\pi^2}{240L^4} \,.
\ee 

\subsection{$\mathcal{N}=4$ super Yang-Mills}
We now consider non-interacting $\mathcal{N}=4$ super Yang-Mills in $3+1$ dimensions. The field content of this theory is a single $U(N)$ gauge field, six adjoint scalars, and four adjoint Weyl fermions. The fermions have antiperiodic boundary conditions along the thermal circle which imply antiperiodic boundary conditions on the spatial $S^1 \times \mathbb{R}^2$  on which the Casimir energy is calculated. The Casimir energy density and thermal entropy at weak coupling were first calculated in \cite{Horowitz:1998ha} and \cite{Gubser:1996de}, respectively. The results are
\be
S=\f{2\pi^2}{3} N^2 V_3 T^3, \qquad \varepsilon_{\textrm{vac}} = \f{\pi^2N^2}{6}\,.
\ee
These are again related by \eqref{formula}.

\section{Summary and outlook}\label{conclusions}
We have used higher-dimensional modular invariance to extend the high-temperature/low-temperature duality of two-dimensional CFTs to higher dimensions. We used the new invariances to derive formulas for the thermal entropy at high temperature and asymptotic degeneracy of states on a torus and provided a few basic checks.  For $d=1$ the formula reduces to the usual Cardy formula. We also provided generalizations to theories which violate hyperscaling and which have an anisotropic scaling symmetry. As a simple application of our formula, we provided a new derivation of the Casimir force between parallel conducting plates in quantum electrodynamics.

The symmetries discussed above may be useful in various applications to entanglement entropy. For example, consider the $n^{\textrm{th}}$ R\'enyi entropy of a disc-shaped region. This can be conformally mapped to the thermal entropy on hyperbolic space \cite{Casini:2011kv}. For $n \ll 1$, the temperature is very large and we can ignore the curvature of the hyperbolic space. In that case we can treat it as a large torus and use our formula \eqref{formula} to obtain the thermal entropy. One can also analyze the second R\'enyi entropy of two strips which are infinite in all directions except one. Under the replica trick this is topologically $\mathbb{T}^2 \times \mathbb{R}^{d-1}$. 

Within holography, these formulas give the microscopic entropy of large AdS-Schwarschild black holes/branes and hyperscaling-violating black branes. Generalizing these formulas to include angular momentum gives the microscopic entropy of boosted black branes and large Kerr-AdS black holes. The strongly coupled vacuum energy of the dual field theory is obtained in all cases from the AdS soliton and its hyperscaling-violating cousins. These applications to holography will be discussed in a forthcoming paper \cite{Shaghoulian:2015lcn}.

\section*{Acknowledgments}
I would like to acknowledge all the people I have asked about these entropy formulas: Dionysios Anninos, John Cardy, Steven Carlip, Aleksey Cherman, Will Donnelly, Gary Horowitz, Daniel Kabat, Raghu Mahajan, David McGady, Joe Polchinski, Joshua Samani, Stephen Shenker, Mark Srednicki, and Douglas Stanford. I would like to especially thank Tarek Anous and Eric Perlmutter for useful discussions, and Dionysios Anninos and Mark Srednicki for comments on a draft. This work is supported by NSF Grant PHY13-16748.

\appendix
\section{Higher-dimensional Cardy formula and near-vacuum states}\label{app}
We have argued that on $\mathbb{T}^2\times \mathbb{T}^{d-1}$, with the cycle lengths $L_\infty$ on the $\mathbb{T}^{d-1}$ much bigger than the cycle lengths on $L, \beta$ on $\mathbb{T}^2$, that we have
\be
\log Z(\beta) = \left(\f{L}{\beta}\right)^{d-1}\log Z\left(\f{L^2}{\beta}\right),
\ee
with corrections suppressed by $L_\infty$. On first glance, the derivation of our formulas \eqref{formula}-\eqref{states} proceeds just as in two dimensions. In particular, one takes $L/\beta$ large  to project to the vacuum state:
\be
Z(\beta)= \sum e^{-\beta E} = \left(\sum e^{-\f{L^2}{\beta} E}\right)^{(L/\beta)^{d-1}}\approx e^{-\f{L^{d+1}}{\beta^d}E_\textrm{vac}}\,.
\ee
However, there is an obvious subtlety. We have $L_\infty\gg L^2/\beta$, so the spacing of states above the vacuum is tiny and there is a degeneracy piling up there. In other words, one may suspect that you cannot cleanly project to the vacuum as in two dimensions. There are different ways to deal with this. One way is to allow $L^2/\beta \gg L_\infty$, even though $L,\beta \ll L_\infty$. This is possible, although it breaks our interpretation of working on a torus with two directions much smaller than the other $d-1$ directions.\footnote{When $L^2/\beta \sim L_\infty$ instead of $L^2/\beta \ll L_\infty$, one may question the argument of high-temperature extensivity to scale out the factor of $(L/\beta)^{d-1}$ to obtain our modular-form structure. However, since we still have $L\ll L_\infty$, quantizing along this direction gives a high-temperature partition function that implies extensivity with respect to the $L_\infty$ directions. By modular invariance, this extensivity extends to the quantization along $L^2/\beta$.}

The other way is to notice that even for $L_\infty \gg L^2/\beta$ the degeneracy of states piling up near the vacuum cannot compete with the leading piece. To illustrate the point, we will consider three dimensions with cycle lengths $\beta \ll L \ll L_\infty$. We will consider quantizations along all three cycles and assume extensivity at high temperature. In that case, we have
\begin{align}
Z(\beta)_{L\times L_\infty} &= \sum e^{-\beta E_{L\times L_\infty}} \approx e^{c L L_\infty/\beta^2},\label{high}\\
Z(L_\infty)_{L\times \beta} &= \sum e^{-L_\infty E_{L\times \beta}} \approx e^{-L_\infty E_{\textrm{vac},\,L\times \beta}}= e^{\varepsilon_{\textrm{vac}} L L_\infty/\beta^2},\label{low}
\end{align}
where $c$ is the thermal coefficient and $\varepsilon_{\textrm{vac}}$ is the number characterizing the vacuum energy on a background that is approximately $S^1 \times \mathbb{R}$. Both quantizations should be equivalent, i.e. $Z(\beta)_{L\times L_\infty}  = Z(L_\infty)_{L\times \beta} $.  As argued in the main text and as apparent from these expressions, $c=\varepsilon_{\textrm{vac}}$. \eqref{high} is just the high-temperature partition function which is fixed by our assumption of extensivity, whereas \eqref{low} is a low-temperature partition function, which projects to the vacuum. In this case $L_\infty\gg L, \beta$ so we have a clean projection. The scaling of the vacuum energy can be argued by matching the result to the quantization along $\beta$. We can now consider quantization along $L$:
\be
Z(L)_{\beta\times L_\infty } = \sum e^{-L E_{\beta \times L_\infty}} =  e^{-L E_{\textrm{vac},\,\beta \times L_\infty}}+\sum_{E>E_{\textrm{vac}}} e^{-L E_{\beta \times L_\infty}}\,.
\ee
In the final form of the expression we have simply split the sum into the contribution of the vacuum state and the contribution from the excited states. Since $L\ll L_\infty$, one may be worried that you cannot only pick up the vacuum contribution and that the sum from the nearly degenerate excited states can contribute. But notice that the vacuum contribution alone gives $e^{\varepsilon_{\textrm{vac}} L L_\infty/\beta^2}$, which agrees with the quantizations along $\beta$ and along $L_\infty$. Thus, the excited states do not compete with the leading term.

These three quantizations also illustrate the connections between corrections to the partition function in the various quantizations. For example, there are sub-extensive contributions to the quantization along $\beta$. These sub-extensive contributions appear as the low-lying excited states that were neglected in the quantization along $L$. So far, this is similar to the story in two spacetime dimensions. However, quantization along $L_\infty$ highlights yet another interesting form of corrections. In this case, using $L_\infty$ as a control parameter, there is no reason to expect the excited states to contribute. Instead, the leading correction comes from the scaling of the vacuum energy on $L\times \beta$, i.e. $E_{\textrm{vac}, \, L\times \beta} = -\varepsilon_{\textrm{vac}}L/\beta^2(1+a_1(\beta/L)+\dots)$ for some constant $a_1$. Such a correction exists in the quantization along $L$ but $\beta/L_\infty \ll \beta/L$ makes it subleading compared to the low-lying excited states. Given such a correction to the ground-state energy, the low-lying excited states must sum up to a contribution in the partition function of the form
\be
Z(L)_{L_\infty \times \beta}= e^{-L E_{\textrm{vac},\,L_\infty\times \beta}}\left(1+\textrm{excited}\right)\approx e^{-L E_{\textrm{vac},\,L_\infty\times \beta}}\left(e^{a_1 \varepsilon_{\textrm{vac}}L_\infty/\beta}+\dots\right)
\ee
to agree with the quantization along $L_\infty$. Of course, the connection of subleading corrections to the free energy corresponding to subleading corrections to the vacuum energy is not a surprise; see e.g. \eqref{vacmass}-\eqref{freemass}. However, it is interesting in this context to see (a) a situation where the subleading corrections to the free energy are sub-extensive (the correction in \eqref{freemass} still scaled with volume), and (b) the equality between the correction due to the low-lying excited states and the correction of the ground state energy itself.
\small{
\bibliographystyle{apsrev4-1long}
\bibliography{cardyhigherCFTbiblio}}

\begin{thebibliography}{10}%
\makeatletter
\providecommand \@ifxundefined [1]{%
 \ifx #1\undefined \expandafter \@firstoftwo
 \else \expandafter \@secondoftwo
\fi
}%
\providecommand \@ifnum [1]{%
 \ifnum #1\expandafter \@firstoftwo
 \else \expandafter \@secondoftwo
\fi
}%
\providecommand \enquote [1]{``#1''}%
\providecommand \bibnamefont  [1]{#1}%
\providecommand \bibfnamefont [1]{#1}%
\providecommand \citenamefont [1]{#1}%
\providecommand\href[0]{\@sanitize\@href}%
\providecommand\@href[1]{\endgroup\@@startlink{#1}\endgroup\@@href}%
\providecommand\@@href[1]{#1\@@endlink}%
\providecommand \@sanitize [0]{\begingroup\catcode`\&12\catcode`\#12\relax}%
\@ifxundefined \pdfoutput {\@firstoftwo}{%
 \@ifnum{\z@=\pdfoutput}{\@firstoftwo}{\@secondoftwo}%
}{%
 \providecommand\@@startlink[1]{\leavevmode\special{html:<a href="#1">}}%
 \providecommand\@@endlink[0]{\special{html:</a>}}%
}{%
 \providecommand\@@startlink[1]{%
  \leavevmode
  \pdfstartlink
   attr{/Border[0 0 1 ]/H/I/C[0 1 1]}%
   user{/Subtype/Link/A<</Type/Action/S/URI/URI(#1)>>}%
  \relax
 }%
 \providecommand\@@endlink[0]{\pdfendlink}%
}%
\providecommand \url  [0]{\begingroup\@sanitize \@url }%
\providecommand \@url [1]{\endgroup\@href {#1}{\urlprefix}}%
\providecommand \urlprefix [0]{URL }%
\providecommand \Eprint[0]{\href }%
\@ifxundefined \urlstyle {%
  \providecommand \doi [1]{doi:\discretionary{}{}{}#1}%
}{%
  \providecommand \doi [0]{doi:\discretionary{}{}{}\begingroup
  \urlstyle{rm}\Url }%
}%
\providecommand \doibase [0]{http://dx.doi.org/}%
\providecommand \Doi[1]{\href{\doibase#1}}%
\providecommand \bibAnnote [3]{%
  \BibitemShut{#1}%
  \begin{quotation}\noindent
    \textsc{Key:}\ #2\\\textsc{Annotation:}\ #3%
  \end{quotation}%
}%
\providecommand \bibAnnoteFile [2]{%
  \IfFileExists{#2}{\bibAnnote {#1} {#2} {\input{#2}}}{}%
}%
\providecommand \typeout [0]{\immediate \write \m@ne }%
\providecommand \selectlanguage [0]{\@gobble}%
\providecommand \bibinfo [0]{\@secondoftwo}%
\providecommand \bibfield [0]{\@secondoftwo}%
\providecommand \translation [1]{[#1]}%
\providecommand \BibitemOpen[0]{}%
\providecommand \bibitemStop [0]{}%
\providecommand \bibitemNoStop [0]{.\EOS\space}%
\providecommand \EOS [0]{\spacefactor3000\relax}%
\providecommand \BibitemShut [1]{\csname bibitem#1\endcsname}%
\bibitem{Kramers:1941kn}%
  \BibitemOpen
  \bibfield{author}{%
  \bibinfo {author} {\bibfnamefont{H.~A.}\ \bibnamefont{Kramers}}\ and\
  \bibinfo {author} {\bibfnamefont{G.~H.}\ \bibnamefont{Wannier}},\ }%
  \bibfield{title}{%
  \enquote{\bibinfo {title} {{Statistics of the two-dimensional ferromagnet.
  Part 1.}}.}\ }%
  \bibfield{journal}{%
  \Doi{10.1103/PhysRev.60.252}{\bibinfo {journal} {Phys. Rev.}}\ }%
  \textbf{\bibinfo {volume} {60}},\ \bibinfo {pages} {252--262} (\bibinfo
  {year} {1941})%
  \bibAnnoteFile{NoStop}{Kramers:1941kn}%
\bibitem{Cardy:1986ie}%
  \BibitemOpen
  \bibfield{author}{%
  \bibinfo {author} {\bibfnamefont{John~L.}\ \bibnamefont{Cardy}},\ }%
  \bibfield{title}{%
  \enquote{\bibinfo {title} {{Operator Content of Two-Dimensional Conformally
  Invariant Theories}},}\ }%
  \bibfield{journal}{%
  \Doi{10.1016/0550-3213(86)90552-3}{\bibinfo {journal} {Nucl.Phys.}}\ }%
  \textbf{\bibinfo {volume} {B270}},\ \bibinfo {pages} {186--204} (\bibinfo
  {year} {1986})%
  \bibAnnoteFile{NoStop}{Cardy:1986ie}%
\bibitem{Cardy:1991kr}%
  \BibitemOpen
  \bibfield{author}{%
  \bibinfo {author} {\bibfnamefont{John~L.}\ \bibnamefont{Cardy}},\ }%
  \bibfield{title}{%
  \enquote{\bibinfo {title} {{Operator content and modular properties of higher
  dimensional conformal field theories}},}\ }%
  \bibfield{journal}{%
  \Doi{10.1016/0550-3213(91)90024-R}{\bibinfo {journal} {Nucl.Phys.}}\ }%
  \textbf{\bibinfo {volume} {B366}},\ \bibinfo {pages} {403--419} (\bibinfo
  {year} {1991})%
  \bibAnnoteFile{NoStop}{Cardy:1991kr}%
\bibitem{Lorenzen:2014pna}%
  \BibitemOpen
  \bibfield{author}{%
  \bibinfo {author} {\bibfnamefont{Jakob}\ \bibnamefont{Lorenzen}}\ and\
  \bibinfo {author} {\bibfnamefont{Dario}\ \bibnamefont{Martelli}},\ }%
  \bibfield{title}{%
  \enquote{\bibinfo {title} {{Comments on the Casimir energy in supersymmetric
  field theories}},}\ }%
  \bibfield{journal}{%
  \Doi{10.1007/JHEP07(2015)001}{\bibinfo {journal} {JHEP}}\ }%
  \textbf{\bibinfo {volume} {07}},\ \bibinfo {pages} {001} (\bibinfo {year}
  {2015}),\ \Eprint{http://arxiv.org/abs/1412.7463}{arXiv:1412.7463 [hep-th]}%
  \bibAnnoteFile{NoStop}{Lorenzen:2014pna}%
\bibitem{Assel:2015nca}%
  \BibitemOpen
  \bibfield{author}{%
  \bibinfo {author} {\bibfnamefont{Benjamin}\ \bibnamefont{Assel}}, \bibinfo
  {author} {\bibfnamefont{Davide}\ \bibnamefont{Cassani}}, \bibinfo {author}
  {\bibfnamefont{Lorenzo}\ \bibnamefont{Di~Pietro}}, \bibinfo {author}
  {\bibfnamefont{Zohar}\ \bibnamefont{Komargodski}}, \bibinfo {author}
  {\bibfnamefont{Jakob}\ \bibnamefont{Lorenzen}}, \emph{et~al.},\ }%
  \bibfield{title}{%
  \enquote{\bibinfo {title} {{The Casimir Energy in Curved Space and its
  Supersymmetric Counterpart}},}\ }%
   (\bibinfo {year} {2015}),\
  \Eprint{http://arxiv.org/abs/1503.05537}{arXiv:1503.05537 [hep-th]}%
  \bibAnnoteFile{NoStop}{Assel:2015nca}%
\bibitem{Verlinde:2000wg}%
  \BibitemOpen
  \bibfield{author}{%
  \bibinfo {author} {\bibfnamefont{Erik~P.}\ \bibnamefont{Verlinde}},\ }%
  \bibfield{title}{%
  \enquote{\bibinfo {title} {{On the holographic principle in a radiation
  dominated universe}},}\ }%
   (\bibinfo {year} {2000}),\
  \Eprint{http://arxiv.org/abs/hep-th/0008140}{arXiv:hep-th/0008140 [hep-th]}%
  \bibAnnoteFile{NoStop}{Verlinde:2000wg}%
\bibitem{Kutasov:2000td}%
  \BibitemOpen
  \bibfield{author}{%
  \bibinfo {author} {\bibfnamefont{David}\ \bibnamefont{Kutasov}}\ and\
  \bibinfo {author} {\bibfnamefont{Finn}\ \bibnamefont{Larsen}},\ }%
  \bibfield{title}{%
  \enquote{\bibinfo {title} {{Partition sums and entropy bounds in weakly
  coupled CFT}},}\ }%
  \bibfield{journal}{%
  \Doi{10.1088/1126-6708/2001/01/001}{\bibinfo {journal} {JHEP}}\ }%
  \textbf{\bibinfo {volume} {0101}},\ \bibinfo {pages} {001} (\bibinfo {year}
  {2001}),\ \Eprint{http://arxiv.org/abs/hep-th/0009244}{arXiv:hep-th/0009244
  [hep-th]}%
  \bibAnnoteFile{NoStop}{Kutasov:2000td}%
\bibitem{DiPietro:2014bca}%
  \BibitemOpen
  \bibfield{author}{%
  \bibinfo {author} {\bibfnamefont{Lorenzo}\ \bibnamefont{Di~Pietro}}\ and\
  \bibinfo {author} {\bibfnamefont{Zohar}\ \bibnamefont{Komargodski}},\ }%
  \bibfield{title}{%
  \enquote{\bibinfo {title} {{Cardy formulae for SUSY theories in $d =$ 4 and
  $d =$ 6}},}\ }%
  \bibfield{journal}{%
  \Doi{10.1007/JHEP12(2014)031}{\bibinfo {journal} {JHEP}}\ }%
  \textbf{\bibinfo {volume} {1412}},\ \bibinfo {pages} {031} (\bibinfo {year}
  {2014}),\ \Eprint{http://arxiv.org/abs/1407.6061}{arXiv:1407.6061 [hep-th]}%
  \bibAnnoteFile{NoStop}{DiPietro:2014bca}%
\bibitem{Ardehali:2015hya}%
  \BibitemOpen
  \bibfield{author}{%
  \bibinfo {author} {\bibfnamefont{Arash~Arabi}\ \bibnamefont{Ardehali}},
  \bibinfo {author} {\bibfnamefont{James~T.}\ \bibnamefont{Liu}},\ and\
  \bibinfo {author} {\bibfnamefont{Phillip}\ \bibnamefont{Szepietowski}},\ }%
  \bibfield{title}{%
  \enquote{\bibinfo {title} {{High-Temperature Expansion of Supersymmetric
  Partition Functions}},}\ }%
  \bibfield{journal}{%
  \Doi{10.1007/JHEP07(2015)113}{\bibinfo {journal} {JHEP}}\ }%
  \textbf{\bibinfo {volume} {07}},\ \bibinfo {pages} {113} (\bibinfo {year}
  {2015}),\ \Eprint{http://arxiv.org/abs/1502.07737}{arXiv:1502.07737
  [hep-th]}%
  \bibAnnoteFile{NoStop}{Ardehali:2015hya}%
\bibitem{Zhou:2015cpa}%
  \BibitemOpen
  \bibfield{author}{%
  \bibinfo {author} {\bibfnamefont{Yang}\ \bibnamefont{Zhou}},\ }%
  \bibfield{title}{%
  \enquote{\bibinfo {title} {{Universal Features of Four-Dimensional
  Superconformal Field Theory on Conic Space}},}\ }%
  \bibfield{journal}{%
  \Doi{10.1007/JHEP08(2015)052}{\bibinfo {journal} {JHEP}}\ }%
  \textbf{\bibinfo {volume} {08}},\ \bibinfo {pages} {052} (\bibinfo {year}
  {2015}),\ \Eprint{http://arxiv.org/abs/1506.06512}{arXiv:1506.06512
  [hep-th]}%
  \bibAnnoteFile{NoStop}{Zhou:2015cpa}%
\bibitem{Landsteiner:2011cp}%
  \BibitemOpen
  \bibfield{author}{%
  \bibinfo {author} {\bibfnamefont{Karl}\ \bibnamefont{Landsteiner}}, \bibinfo
  {author} {\bibfnamefont{Eugenio}\ \bibnamefont{Megias}},\ and\ \bibinfo
  {author} {\bibfnamefont{Francisco}\ \bibnamefont{Pena-Benitez}},\ }%
  \bibfield{title}{%
  \enquote{\bibinfo {title} {{Gravitational Anomaly and Transport}},}\ }%
  \bibfield{journal}{%
  \Doi{10.1103/PhysRevLett.107.021601}{\bibinfo {journal} {Phys. Rev. Lett.}}\
  }%
  \textbf{\bibinfo {volume} {107}},\ \bibinfo {pages} {021601} (\bibinfo {year}
  {2011}),\ \Eprint{http://arxiv.org/abs/1103.5006}{arXiv:1103.5006 [hep-ph]}%
  \bibAnnoteFile{NoStop}{Landsteiner:2011cp}%
\bibitem{Landsteiner:2011iq}%
  \BibitemOpen
  \bibfield{author}{%
  \bibinfo {author} {\bibfnamefont{Karl}\ \bibnamefont{Landsteiner}}, \bibinfo
  {author} {\bibfnamefont{Eugenio}\ \bibnamefont{Megias}}, \bibinfo {author}
  {\bibfnamefont{Luis}\ \bibnamefont{Melgar}},\ and\ \bibinfo {author}
  {\bibfnamefont{Francisco}\ \bibnamefont{Pena-Benitez}},\ }%
  \bibfield{title}{%
  \enquote{\bibinfo {title} {{Holographic Gravitational Anomaly and Chiral
  Vortical Effect}},}\ }%
  \bibfield{journal}{%
  \Doi{10.1007/JHEP09(2011)121}{\bibinfo {journal} {JHEP}}\ }%
  \textbf{\bibinfo {volume} {09}},\ \bibinfo {pages} {121} (\bibinfo {year}
  {2011}),\ \Eprint{http://arxiv.org/abs/1107.0368}{arXiv:1107.0368 [hep-th]}%
  \bibAnnoteFile{NoStop}{Landsteiner:2011iq}%
\bibitem{Loganayagam:2012pz}%
  \BibitemOpen
  \bibfield{author}{%
  \bibinfo {author} {\bibfnamefont{R.}~\bibnamefont{Loganayagam}}\ and\
  \bibinfo {author} {\bibfnamefont{Piotr}\ \bibnamefont{Surowka}},\ }%
  \bibfield{title}{%
  \enquote{\bibinfo {title} {{Anomaly/Transport in an Ideal Weyl gas}},}\ }%
  \bibfield{journal}{%
  \Doi{10.1007/JHEP04(2012)097}{\bibinfo {journal} {JHEP}}\ }%
  \textbf{\bibinfo {volume} {04}},\ \bibinfo {pages} {097} (\bibinfo {year}
  {2012}),\ \Eprint{http://arxiv.org/abs/1201.2812}{arXiv:1201.2812 [hep-th]}%
  \bibAnnoteFile{NoStop}{Loganayagam:2012pz}%
\bibitem{Loganayagam:2012zg}%
  \BibitemOpen
  \bibfield{author}{%
  \bibinfo {author} {\bibfnamefont{R.}~\bibnamefont{Loganayagam}},\ }%
  \bibfield{title}{%
  \enquote{\bibinfo {title} {{Anomalies and the Helicity of the Thermal
  State}},}\ }%
  \bibfield{journal}{%
  \Doi{10.1007/JHEP11(2013)205}{\bibinfo {journal} {JHEP}}\ }%
  \textbf{\bibinfo {volume} {11}},\ \bibinfo {pages} {205} (\bibinfo {year}
  {2013}),\ \Eprint{http://arxiv.org/abs/1211.3850}{arXiv:1211.3850 [hep-th]}%
  \bibAnnoteFile{NoStop}{Loganayagam:2012zg}%
\bibitem{Jensen:2012kj}%
  \BibitemOpen
  \bibfield{author}{%
  \bibinfo {author} {\bibfnamefont{Kristan}\ \bibnamefont{Jensen}}, \bibinfo
  {author} {\bibfnamefont{R.}~\bibnamefont{Loganayagam}},\ and\ \bibinfo
  {author} {\bibfnamefont{Amos}\ \bibnamefont{Yarom}},\ }%
  \bibfield{title}{%
  \enquote{\bibinfo {title} {{Thermodynamics, gravitational anomalies and
  cones}},}\ }%
  \bibfield{journal}{%
  \Doi{10.1007/JHEP02(2013)088}{\bibinfo {journal} {JHEP}}\ }%
  \textbf{\bibinfo {volume} {02}},\ \bibinfo {pages} {088} (\bibinfo {year}
  {2013}),\ \Eprint{http://arxiv.org/abs/1207.5824}{arXiv:1207.5824 [hep-th]}%
  \bibAnnoteFile{NoStop}{Jensen:2012kj}%
\bibitem{Jensen:2013kka}%
  \BibitemOpen
  \bibfield{author}{%
  \bibinfo {author} {\bibfnamefont{Kristan}\ \bibnamefont{Jensen}}, \bibinfo
  {author} {\bibfnamefont{R.}~\bibnamefont{Loganayagam}},\ and\ \bibinfo
  {author} {\bibfnamefont{Amos}\ \bibnamefont{Yarom}},\ }%
  \bibfield{title}{%
  \enquote{\bibinfo {title} {{Anomaly inflow and thermal equilibrium}},}\ }%
  \bibfield{journal}{%
  \Doi{10.1007/JHEP05(2014)134}{\bibinfo {journal} {JHEP}}\ }%
  \textbf{\bibinfo {volume} {05}},\ \bibinfo {pages} {134} (\bibinfo {year}
  {2014}),\ \Eprint{http://arxiv.org/abs/1310.7024}{arXiv:1310.7024 [hep-th]}%
  \bibAnnoteFile{NoStop}{Jensen:2013kka}%
\bibitem{Jensen:2013rga}%
  \BibitemOpen
  \bibfield{author}{%
  \bibinfo {author} {\bibfnamefont{Kristan}\ \bibnamefont{Jensen}}, \bibinfo
  {author} {\bibfnamefont{R.}~\bibnamefont{Loganayagam}},\ and\ \bibinfo
  {author} {\bibfnamefont{Amos}\ \bibnamefont{Yarom}},\ }%
  \bibfield{title}{%
  \enquote{\bibinfo {title} {{Chern-Simons terms from thermal circles and
  anomalies}},}\ }%
  \bibfield{journal}{%
  \Doi{10.1007/JHEP05(2014)110}{\bibinfo {journal} {JHEP}}\ }%
  \textbf{\bibinfo {volume} {05}},\ \bibinfo {pages} {110} (\bibinfo {year}
  {2014}),\ \Eprint{http://arxiv.org/abs/1311.2935}{arXiv:1311.2935 [hep-th]}%
  \bibAnnoteFile{NoStop}{Jensen:2013rga}%
\bibitem{Azeyanagi:2015gqa}%
  \BibitemOpen
  \bibfield{author}{%
  \bibinfo {author} {\bibfnamefont{Tatsuo}\ \bibnamefont{Azeyanagi}}, \bibinfo
  {author} {\bibfnamefont{R.}~\bibnamefont{Loganayagam}},\ and\ \bibinfo
  {author} {\bibfnamefont{Gim~Seng}\ \bibnamefont{Ng}},\ }%
  \bibfield{title}{%
  \enquote{\bibinfo {title} {{Anomalies, Chern-Simons Terms and Black Hole
  Entropy}},}\ }%
   (\bibinfo {year} {2015}),\
  \Eprint{http://arxiv.org/abs/1505.02816}{arXiv:1505.02816 [hep-th]}%
  \bibAnnoteFile{NoStop}{Azeyanagi:2015gqa}%
\bibitem{Detournay:2012pc}%
  \BibitemOpen
  \bibfield{author}{%
  \bibinfo {author} {\bibfnamefont{Stephane}\ \bibnamefont{Detournay}},
  \bibinfo {author} {\bibfnamefont{Thomas}\ \bibnamefont{Hartman}},\ and\
  \bibinfo {author} {\bibfnamefont{Diego~M.}\ \bibnamefont{Hofman}},\ }%
  \bibfield{title}{%
  \enquote{\bibinfo {title} {{Warped Conformal Field Theory}},}\ }%
  \bibfield{journal}{%
  \Doi{10.1103/PhysRevD.86.124018}{\bibinfo {journal} {Phys. Rev.}}\ }%
  \textbf{\bibinfo {volume} {D86}},\ \bibinfo {pages} {124018} (\bibinfo {year}
  {2012}),\ \Eprint{http://arxiv.org/abs/1210.0539}{arXiv:1210.0539 [hep-th]}%
  \bibAnnoteFile{NoStop}{Detournay:2012pc}%
\bibitem{Bagchi:2012xr}%
  \BibitemOpen
  \bibfield{author}{%
  \bibinfo {author} {\bibfnamefont{Arjun}\ \bibnamefont{Bagchi}}, \bibinfo
  {author} {\bibfnamefont{Stéphane}\ \bibnamefont{Detournay}}, \bibinfo
  {author} {\bibfnamefont{Reza}\ \bibnamefont{Fareghbal}},\ and\ \bibinfo
  {author} {\bibfnamefont{Joan}\ \bibnamefont{Simón}},\ }%
  \bibfield{title}{%
  \enquote{\bibinfo {title} {{Holography of 3D Flat Cosmological Horizons}},}\
  }%
  \bibfield{journal}{%
  \Doi{10.1103/PhysRevLett.110.141302}{\bibinfo {journal} {Phys. Rev. Lett.}}\
  }%
  \textbf{\bibinfo {volume} {110}},\ \bibinfo {pages} {141302} (\bibinfo {year}
  {2013}),\ \Eprint{http://arxiv.org/abs/1208.4372}{arXiv:1208.4372 [hep-th]}%
  \bibAnnoteFile{NoStop}{Bagchi:2012xr}%
\bibitem{Shaghoulian:2015dwa}%
  \BibitemOpen
  \bibfield{author}{%
  \bibinfo {author} {\bibfnamefont{Edgar}\ \bibnamefont{Shaghoulian}},\ }%
  \bibfield{title}{%
  \enquote{\bibinfo {title} {{A Cardy formula for holographic
  hyperscaling-violating theories}},}\ }%
  \bibfield{journal}{%
  \Doi{10.1007/JHEP11(2015)081}{\bibinfo {journal} {JHEP}}\ }%
  \textbf{\bibinfo {volume} {11}},\ \bibinfo {pages} {081} (\bibinfo {year}
  {2015}),\ \Eprint{http://arxiv.org/abs/1504.02094}{arXiv:1504.02094
  [hep-th]}%
  \bibAnnoteFile{NoStop}{Shaghoulian:2015dwa}%
\bibitem{Bravo-Gaete:2015wua}%
  \BibitemOpen
  \bibfield{author}{%
  \bibinfo {author} {\bibfnamefont{Moises}\ \bibnamefont{Bravo-Gaete}},
  \bibinfo {author} {\bibfnamefont{Sebastian}\ \bibnamefont{Gomez}},\ and\
  \bibinfo {author} {\bibfnamefont{Mokhtar}\ \bibnamefont{Hassaine}},\ }%
  \bibfield{title}{%
  \enquote{\bibinfo {title} {{Towards the Cardy formula for hyperscaling
  violation black holes}},}\ }%
  \bibfield{journal}{%
  \Doi{10.1103/PhysRevD.91.124038}{\bibinfo {journal} {Phys. Rev.}}\ }%
  \textbf{\bibinfo {volume} {D91}},\ \bibinfo {pages} {124038} (\bibinfo {year}
  {2015}),\ \Eprint{http://arxiv.org/abs/1505.00702}{arXiv:1505.00702
  [hep-th]}%
  \bibAnnoteFile{NoStop}{Bravo-Gaete:2015wua}%
\bibitem{Cappelli:1988vw}%
  \BibitemOpen
  \bibfield{author}{%
  \bibinfo {author} {\bibfnamefont{Andrea}\ \bibnamefont{Cappelli}}\ and\
  \bibinfo {author} {\bibfnamefont{Antoine}\ \bibnamefont{Coste}},\ }%
  \bibfield{title}{%
  \enquote{\bibinfo {title} {{On the Stress Tensor of Conformal Field Theories
  in Higher Dimensions}},}\ }%
  \bibfield{journal}{%
  \Doi{10.1016/0550-3213(89)90414-8}{\bibinfo {journal} {Nucl.Phys.}}\ }%
  \textbf{\bibinfo {volume} {B314}},\ \bibinfo {pages} {707} (\bibinfo {year}
  {1989})%
  \bibAnnoteFile{NoStop}{Cappelli:1988vw}%
\bibitem{Solodukhin:2011gn}%
  \BibitemOpen
  \bibfield{author}{%
  \bibinfo {author} {\bibfnamefont{Sergey~N.}\ \bibnamefont{Solodukhin}},\ }%
  \bibfield{title}{%
  \enquote{\bibinfo {title} {{Entanglement entropy of black holes}},}\ }%
  \bibfield{journal}{%
  \bibinfo {journal} {Living Rev.Rel.}\ }%
  \textbf{\bibinfo {volume} {14}},\ \bibinfo {pages} {8} (\bibinfo {year}
  {2011}),\ \Eprint{http://arxiv.org/abs/1104.3712}{arXiv:1104.3712 [hep-th]}%
  \bibAnnoteFile{NoStop}{Solodukhin:2011gn}%
\bibitem{Ambjorn:1981xw}%
  \BibitemOpen
  \bibfield{author}{%
  \bibinfo {author} {\bibfnamefont{Jan}\ \bibnamefont{Ambjorn}}\ and\ \bibinfo
  {author} {\bibfnamefont{Stephen}\ \bibnamefont{Wolfram}},\ }%
  \bibfield{title}{%
  \enquote{\bibinfo {title} {{Properties of the Vacuum. 1. Mechanical and
  Thermodynamic}},}\ }%
  \bibfield{journal}{%
  \Doi{10.1016/0003-4916(83)90065-9}{\bibinfo {journal} {Annals Phys.}}\ }%
  \textbf{\bibinfo {volume} {147}},\ \bibinfo {pages} {1} (\bibinfo {year}
  {1983})%
  \bibAnnoteFile{NoStop}{Ambjorn:1981xw}%
\bibitem{Casimir:1948dh}%
  \BibitemOpen
  \bibfield{author}{%
  \bibinfo {author} {\bibfnamefont{H.B.G.}\ \bibnamefont{Casimir}},\ }%
  \bibfield{title}{%
  \enquote{\bibinfo {title} {{On the Attraction Between Two Perfectly
  Conducting Plates}},}\ }%
  \bibfield{journal}{%
  \bibinfo {journal} {Indag.Math.}\ }%
  \textbf{\bibinfo {volume} {10}},\ \bibinfo {pages} {261--263} (\bibinfo
  {year} {1948})%
  \bibAnnoteFile{NoStop}{Casimir:1948dh}%
\bibitem{Brown:1969na}%
  \BibitemOpen
  \bibfield{author}{%
  \bibinfo {author} {\bibfnamefont{Lowell~S.}\ \bibnamefont{Brown}}\ and\
  \bibinfo {author} {\bibfnamefont{G.~Jordan}\ \bibnamefont{Maclay}},\ }%
  \bibfield{title}{%
  \enquote{\bibinfo {title} {{Vacuum stress between conducting plates: An Image
  solution}},}\ }%
  \bibfield{journal}{%
  \Doi{10.1103/PhysRev.184.1272}{\bibinfo {journal} {Phys.Rev.}}\ }%
  \textbf{\bibinfo {volume} {184}},\ \bibinfo {pages} {1272--1279} (\bibinfo
  {year} {1969})%
  \bibAnnoteFile{NoStop}{Brown:1969na}%
\bibitem{Horowitz:1998ha}%
  \BibitemOpen
  \bibfield{author}{%
  \bibinfo {author} {\bibfnamefont{Gary~T.}\ \bibnamefont{Horowitz}}\ and\
  \bibinfo {author} {\bibfnamefont{Robert~C.}\ \bibnamefont{Myers}},\ }%
  \bibfield{title}{%
  \enquote{\bibinfo {title} {{The AdS / CFT correspondence and a new positive
  energy conjecture for general relativity}},}\ }%
  \bibfield{journal}{%
  \Doi{10.1103/PhysRevD.59.026005}{\bibinfo {journal} {Phys.Rev.}}\ }%
  \textbf{\bibinfo {volume} {D59}},\ \bibinfo {pages} {026005} (\bibinfo {year}
  {1998}),\ \Eprint{http://arxiv.org/abs/hep-th/9808079}{arXiv:hep-th/9808079
  [hep-th]}%
  \bibAnnoteFile{NoStop}{Horowitz:1998ha}%
\bibitem{Gubser:1996de}%
  \BibitemOpen
  \bibfield{author}{%
  \bibinfo {author} {\bibfnamefont{S.S.}\ \bibnamefont{Gubser}}, \bibinfo
  {author} {\bibfnamefont{Igor~R.}\ \bibnamefont{Klebanov}},\ and\ \bibinfo
  {author} {\bibfnamefont{A.W.}\ \bibnamefont{Peet}},\ }%
  \bibfield{title}{%
  \enquote{\bibinfo {title} {{Entropy and temperature of black 3-branes}},}\ }%
  \bibfield{journal}{%
  \Doi{10.1103/PhysRevD.54.3915}{\bibinfo {journal} {Phys.Rev.}}\ }%
  \textbf{\bibinfo {volume} {D54}},\ \bibinfo {pages} {3915--3919} (\bibinfo
  {year} {1996}),\
  \Eprint{http://arxiv.org/abs/hep-th/9602135}{arXiv:hep-th/9602135 [hep-th]}%
  \bibAnnoteFile{NoStop}{Gubser:1996de}%
\bibitem{Casini:2011kv}%
  \BibitemOpen
  \bibfield{author}{%
  \bibinfo {author} {\bibfnamefont{Horacio}\ \bibnamefont{Casini}}, \bibinfo
  {author} {\bibfnamefont{Marina}\ \bibnamefont{Huerta}},\ and\ \bibinfo
  {author} {\bibfnamefont{Robert~C.}\ \bibnamefont{Myers}},\ }%
  \bibfield{title}{%
  \enquote{\bibinfo {title} {{Towards a derivation of holographic entanglement
  entropy}},}\ }%
  \bibfield{journal}{%
  \Doi{10.1007/JHEP05(2011)036}{\bibinfo {journal} {JHEP}}\ }%
  \textbf{\bibinfo {volume} {1105}},\ \bibinfo {pages} {036} (\bibinfo {year}
  {2011}),\ \Eprint{http://arxiv.org/abs/1102.0440}{arXiv:1102.0440 [hep-th]}%
  \bibAnnoteFile{NoStop}{Casini:2011kv}%
\bibitem{Shaghoulian:2015lcn}%
  \BibitemOpen
  \bibfield{author}{%
  \bibinfo {author} {\bibfnamefont{Edgar}\ \bibnamefont{Shaghoulian}},\ }%
  \bibfield{title}{%
  \enquote{\bibinfo {title} {{Black hole microstates in AdS}},}\ }%
   (\bibinfo {year} {2015}),\
  \Eprint{http://arxiv.org/abs/1512.06855}{arXiv:1512.06855 [hep-th]}%
  \bibAnnoteFile{NoStop}{Shaghoulian:2015lcn}%
\end{thebibliography}%

\end{document}